\documentclass[aps,prl,twocolumn,superscriptaddress]{revtex4}
\usepackage{graphicx}
\usepackage{color}

\definecolor{orange}{rgb}{1,0.5,0}

\begin{document}

\title{Spin and Valley Control of Free Carriers in Single-Layer WS$_2$}
\author{S{\o}ren~Ulstrup}
\affiliation{Advanced Light Source, E. O. Lawrence Berkeley National Laboratory, Berkeley,
California 94720, USA}
\author{Antonija Grubi\v{s}i\'{c} \v{C}abo}
\affiliation{Department of Physics and Astronomy, Interdisciplinary Nanoscience Center, Aarhus University,
8000 Aarhus C, Denmark}
\author{Deepnarayan Biswas}
\author{Jonathon M. Riley}
\affiliation{SUPA, School of Physics and Astronomy, University of St. Andrews,
St. Andrews, United Kingdom}
\author{Maciej Dendzik}
\author{Charlotte E. Sanders}
\author{Marco Bianchi}
\affiliation{Department of Physics and Astronomy, Interdisciplinary Nanoscience Center, Aarhus University,
8000 Aarhus C, Denmark}
\author{Cephise Cacho}
\author{Dan Matselyukh}
\author{Richard T. Chapman}
\author{Emma Springate}
\affiliation{Central Laser Facility, STFC Rutherford Appleton Laboratory, Harwell, United Kingdom}
\author{Phil D. C. King}
\affiliation{SUPA, School of Physics and Astronomy, University of St. Andrews,
St. Andrews, United Kingdom}
\author{Jill A. Miwa}
\author{Philip~Hofmann}
\affiliation{Department of Physics and Astronomy, Interdisciplinary Nanoscience Center, Aarhus University,
8000 Aarhus C, Denmark}
\email{philip@phys.au.dk}
\date{\today}

\begin{abstract}
The semiconducting single-layer transition metal dichalcogenides have been identified as ideal materials for accessing and manipulating spin- and valley-quantum numbers due to a set of favorable optical selection rules in these materials. Here, we apply time- and angle-resolved photoemission spectroscopy to directly probe optically excited free carriers in the electronic band structure of a high quality single layer of WS$_2$. We observe that the optically generated free hole density in a single valley can be increased by a factor of 2 using a circularly polarized optical excitation. Moreover, we find that by varying the photon energy of the excitation we can tune the free carrier density in a given spin-split state around the valence band maximum of the material. The control of the photon energy and polarization of the excitation thus permits us to selectively excite free electron-hole pairs with a given spin and within a single valley.  
\end{abstract}
\maketitle

In two-dimensional (2D) materials, control of the spin- and valley-degrees of freedom has been suggested as a new type of tuning knob for carrier dynamics \cite{Rycerz:2007}. Single-layer (SL) transition metal dichalcogenides (TMDCs) such as SL MoS$_2$ and WS$_2$ are particularly promising candidates for new spin- and valley-tronic device paradigms, owing to the broken inversion symmetry of their crystal lattices, a strong spin-orbit coupling and a direct band gap at the $\bar{K}$ and $\bar{K}^{\prime}$ valleys in the electronic structure of the materials \cite{xiaocoupled2012,makatomically2010,wang2012}. These properties have been indirectly verified in SL TMDCs through selective excitation of bound electron-hole pairs with circularly polarized light, leading to the observation of valley polarized excitons \cite{zengvalley2012,makcontrol2012,cao2012,wangvalley2013}, which could be coherently manipulated \cite{Jones:2013}. Device measurements have shown indications of spin- and valley-coupled photocurrents \cite{Yuan:2014} and Hall effects \cite{Mak:2014,Xu:2014c}. Direct measurements of the electronic structure and associated quasiparticles have been carried out for the bulk TMDC material WSe$_2$ using photoemission spectroscopies with spin \cite{Riley:2014} and time resolution \cite{Bertoni:2016}. Such measurements can, to some degree, even give information about the situation in a single layer, due to the surface sensitivity of photoemission spectroscopy. However, there is a need for experimental evidence and quantification of these properties for free carriers in the electronic structure of actual SL TMDCs, which are truly non-inversion symmetric materials.

Time- and angle-resolved photoemission spectroscopy (TR-ARPES) is a powerful experimental approach for detecting optically excited free carriers with  time-, energy- and momentum-resolution with extremely high sensitivity towards 2D materials \cite{Johannsen:2013aa,Gierz:2013aa}. In this technique, a pump pulse with tuneable photon energy and polarization optically excites the material in question. The excited state is then probed by ARPES with an ultraviolet pulse, produced via high harmonic generation (HHG), which can provide sufficiently high photon energies to reach the $\bar{K}$ valleys at the corner of the SL TMDC's Brillouin zone (BZ). Since the two pulses are time-delayed with a resolution on the order of 30~fs, it is possible to stroboscopically record the ultrafast dynamics in the band structure of the material \cite{Perfetti:2007,Rohwer:2011}. In the context of SL TMDCs, such experiments have focused on the size of the screening- and free carrier-induced quasiparticle gap renormalization on various types of substrates \cite{Antonija-Grubisic-Cabo:2015aa,Ulstrup:2016}. A SL TMDC with strong spin-orbit coupling is desirable for a TR-ARPES experiment that endeavors to directly detect free carriers selectively excited in the spin-split states in a given valley \cite{Ulstrup:2014ac}. We therefore use epitaxial SL WS$_2$ grown on Ag(111). WS$_2$ is characterized by a strong spin-splitting on the order of 420~meV at the valence band maximum (VBM) \cite{Dendzik2015}.  Intriguingly, our experiments reveal excited electron and hole populations that are strongly peaked for a resonant excitation between the upper VB spin state and the conduction band (CB). A significant valley polarization of free carriers develops in these conditions when we pump the material with circularly polarized light.

\begin{figure*}
\includegraphics[width=0.75\textwidth]{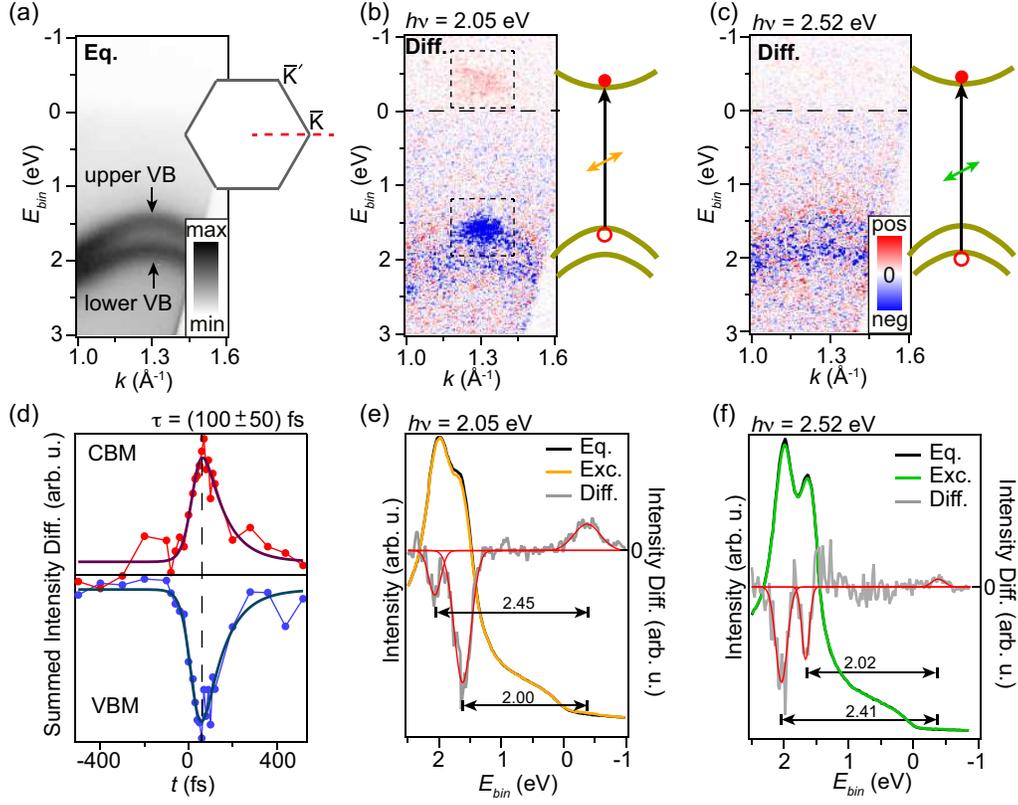}\\
\caption{(Color online) (a) ARPES spectrum of WS$_2$ taken before optical excitation. The BZ in the inset shows the measurement direction via the dashed line. (b)-(c) Difference spectra after and before excitation obtained with an $s$-polarized pump pulse at a photon energy of (b) 2.05~eV and (c) 2.52~eV. The sketches in (b)-(c) present the possible transitions between the VB and CB for the given pump pulse photon energies. (d) Time dependent intensity difference (markers) summed over the dashed boxed regions on the CBM and VBM, respectively, in (b). Overlaid thick lines are fits to a function consisting of an exponential rising edge and a single exponential decay with time constant $\tau$. (e)-(f) Integrated EDCs at equilibrium (Eq.) and peak excitation (Exc.) along with their difference (Diff.) for the corresponding data in (b)-(c). The integration was carried out over the VBM and CBM in the momentum range from 1.15~\AA$^{-1}$ to 1.45~\AA$^{-1}$. Gaussian fits of the peaks in the difference are shown along with estimates of the transition energies in eV (error bars are $\pm 0.05$~eV).}
  \label{fig:1}
\end{figure*}

SL WS$_2$ at a coverage of approximately 0.7~monolayers was grown on a clean Ag(111) surface in ultra-high vacuum (UHV) by deposition of W atoms and annealing to 875~K in a background pressure of 2.5$\times$10$^{-5}$~mbar of H$_2$S gas. The growth on Ag(111) is similar to the growth described for SL WS$_2$ on Au(111) in Ref. \cite{Dendzik2015}. The sample was transported to the Artemis TR-ARPES end-station at the Central Laser Facility at the Harwell Campus in the United Kingdom \cite{Frassetto:2011} in an evacuated tube and cleaned by annealing to 500~K in the UHV chamber. Pump and probe pulses for TR-ARPES were generated by a 1~kHz Ti:sapphire laser. A $p$-polarized probe pulse was used for HHG in an Ar gas cell. We used a harmonic with a photon energy of 25~eV for ARPES. A tuneable pump pulse was achieved by seeding an optical parametric amplifier (TOPAS), enabling optical excitation energies (wavelengths) of 2.05~eV (605 nm) and 2.52~eV (492 nm). The polarization of the pump pulse was controlled with a quarter waveplate, which could be rotated with a motor in order to switch between $s$-, $\sigma^{-}$- and $\sigma^{+}$-polarization, permitting measurements with all three polarizations. The fluence of the pump pulse was kept around 3.0~mJ/cm$^2$ in order to obtain maximum pump-probe signal while avoiding space-charge effects \cite{Ulstrup:2015j}. The average acquisition time for pump-probe data was 24~hours. We used variable exit slit settings for the probe pulse and entrance slit sizes of the SPECS Phoibos 100 electron analyzer depending on the statistics required for the measurement, such that the energy resolution varies between 150-350~meV. The angular and time resolution were on the order of 0.2$^{\circ}$ and 40~fs, respectively. The temperature of the sample was kept at 70~K during measurements. 

In Fig. \ref{fig:1}(a) we present the electronic structure of our SL WS$_2$ sample around the $\bar{K}$-valley, measured by ARPES along the direction of the BZ shown in the inset. The spectrum was obtained in equilibrium conditions before the arrival of the optical excitation. Clearly resolved spin-split bands of the VBM are observed in the spectrum. Throughout the paper we refer to the band on the lower (higher) binding energy side as the upper (lower) VB.  The effect of pumping can be best seen in the difference between the spectrum in the excited state and the equilibrium spectrum in Fig. \ref{fig:1}(a). This difference is taken at a time delay corresponding to the peak of the excitation ($t= 30$~fs). Excitation with an $s$-polarized pump pulse with a photon energy of 2.05~eV leads to a population of excited holes and electrons in the upper VB and conduction band minimum (CBM), which is seen via the blue and red contrast, respectively, in the difference spectrum in Fig. \ref{fig:1}(b). Tuning the pump pulse photon energy to 2.52~eV results in a significantly reduced signal from excited electrons and holes, as shown in Fig. \ref{fig:1}(c), which we discuss in further detail in connection with Fig. 4. The time dependence of the observed difference signals reveals a symmetric and ultrafast decay of excited carriers in the CBM and VBM regions with a time constant $\tau$ of 100~fs, as shown in Fig. \ref{fig:1}(d). These dynamics are found to be independent of the pump pulse energy within our time resolution. This time-dependence reflects the coupling of the SL WS$_2$ to the underlying metallic substrate, which efficiently drains optically excited free carriers, as previously observed for SL MoS$_2$ on Au(111) \cite{Antonija-Grubisic-Cabo:2015aa}. Since the time-dependence of the carrier relaxation is dominated by the coupling to the substrate, we will focus on the conditions of the carrier excitation process. 

\begin{figure}
\includegraphics[width=0.49\textwidth]{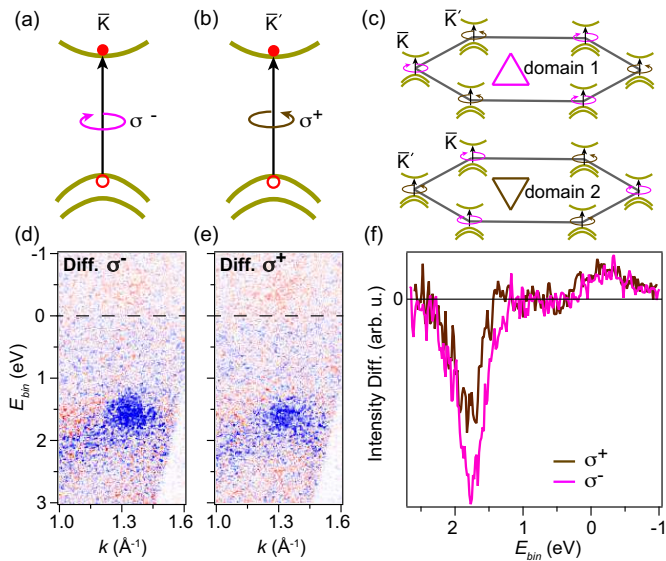}\\
\caption{(Color online) (a)-(b) Selection rules for transitions from the upper VB to the CBM in the $\bar{K}$ and $\bar{K}^{\prime}$ valleys. (c) BZ orientations and selection rules for the two possible domain orientations probed in this work. (d)-(e) Difference signal for optical pumping with (d) $\sigma^{-}$ and (e) $\sigma^{+}$ circular polarization and a photon energy of 2.05~eV. (f) EDCs of the difference integrated over the VBM and CBM regions for the data in (d)-(e).}
  \label{fig:2}
\end{figure}

The analysis in Figs. \ref{fig:1}(e) and \ref{fig:1}(f) presents the integrated energy distribution curves (EDCs) for the equilibrium and excited state spectra along with their difference. Gaussian fits of the peaks in the difference reveal how the upper and lower VBs and the CBM line up. In Fig. \ref{fig:1}(e) we find a direct quasiparticle gap at $\bar{K}$ of (2.00 $\pm$ 0.05)~eV, such that our choice of pump pulse energy of 2.05~eV for this data set is essentially resonant with the transition between the upper VB and the CBM. Fig. \ref{fig:1}(f) reveals a stronger depletion of intensity in the lower VB and a gap of (2.41 $\pm$ 0.05)~eV between the lower VB and the CBM, which is nearly resonant with our pump pulse energy of 2.52~eV in this data set. The quasiparticle gap energy is somewhat smaller than the average value of 2.43~eV calculated for freestanding SL WS$_2$ in the GW approximation \cite{Rasmussen:2015aa}, which is fully consistent with a screening-induced band gap renormalization by the metal substrate \cite{Antonija-Grubisic-Cabo:2015aa,Bruix:2016}. 

\begin{figure}
\includegraphics[width=0.49\textwidth]{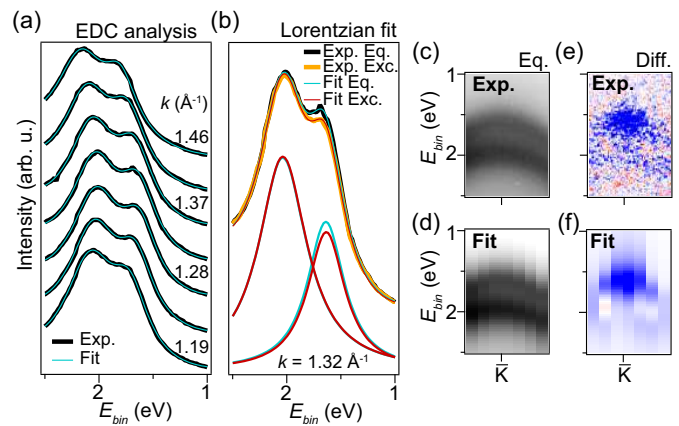}\\
\caption{(Color online) (a) EDCs extracted over the VBM and fitted with two Lorentzian line shapes on a parabolic background. Each EDC is binned over a range of $\pm0.02$~\AA$^{-1}$. (b) Example EDC fit for equilibrium (Eq.) and optically excited (Exc.) data with plots of the individual Lorentzian components. (c) Measured spectrum used for the EDC analysis in (a). (d) Reconstruction of the spectrum using the fitted EDCs. (e)-(f) Difference (Diff.) spectra corresponding to the data and fit in (c)-(d). The spectrum analyzed here was measured with an $s$-polarized optical excitation at a photon energy of 2.05~eV (see also Fig. \ref{fig:1}(b)).}
  \label{fig:3}
\end{figure}

Figs. \ref{fig:2}(a)-(b) sketch the selection rules for optical pumping of SL WS$_2$ with circularly polarized pulses with a photon energy of 2.05~eV  \cite{xiaocoupled2012}. For $\sigma^{-}$-polarized pump pulses, carriers are exclusively excited in the upper VB at $\bar{K}$ (see Fig. \ref{fig:2}(a)), while for $\sigma^{+}$-polarized pump pulses carriers are exclusively excited in the upper VB at $\bar{K}^{\prime}$ (see Fig. \ref{fig:2}(b)). In the present case of an epitaxial WS$_2$ layer growing on Ag(111), the substrate symmetry permits the existence of mirror domains \cite{Lehtinen:2015aa}, implying that we are measuring a superposition of the two BZs with the valleys swapped, as shown in Fig. \ref{fig:2}(c). Regardless of these domains, our experiment reveals a stronger depletion of the upper VB intensity when we use pump pulses with $\sigma^{-}$-polarization. This is seen via the comparison of the intensity difference for the excited holes in the VB generated by $\sigma^{-}$- and $\sigma^{+}$-polarized pulses in Figs. \ref{fig:2}(d) and \ref{fig:2}(e) and the integrated EDCs of the difference for both polarizations in Fig. \ref{fig:2}(f). Note that the existence of mirror domains does not necessarily imply an equal distribution of domain types and even single domain orientations can be achieved in similar systems, depending on the growth conditions \cite{Orlando:2014aa}. Based on the observed circular dichroism here, we conclude that the domain distribution is not equal in the present case. 

In order to obtain a quantitative estimate of the optically induced free hole density we proceed with a band-resolved line shape analysis of the VBM region. EDCs around the VBM of the spectra obtained with an optical excitation energy of 2.05~eV and $s$-polarized light are fitted with two Lorentzian line shapes on a parabolic background, as shown in Fig. \ref{fig:3}(a). The fitted Lorentzian components of the spin-split VBs are extracted as shown in Fig. \ref{fig:3}(b), and we observe a good agreement between the experimental dispersion in Fig. \ref{fig:3}(c) and the fitted dispersion in Fig. \ref{fig:3}(d). The fit captures position and amplitude changes in the line shape, which could occur due to gap renormalization and optically excited holes, respectively. Comparisons of the Lorentzian fits in Fig. \ref{fig:3}(b) and the difference spectra in Figs. \ref{fig:3}(e) and \ref{fig:3}(f) show that the main change in the excited spectrum relates to an amplitude change in the upper VB. This can be translated into a number for the excited holes, $N_h$, per momentum state by integration of the fitted Lorentzian components, normalization to the spectral weight in the equilibrium spectrum, and finally conversion of the normalized spectral weight in the excited spectrum to a value for $N_h$ \cite{Ulstrup:2016}.  

\begin{figure}[t!]
 \includegraphics[width=0.49\textwidth]{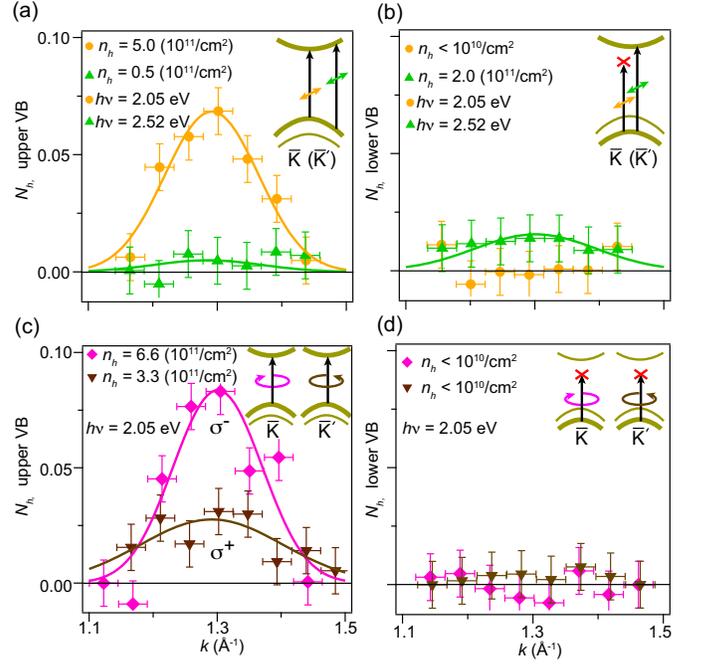}\\
  \caption{(Color online) (a)-(d) Number of holes $N_h$ per momentum state (markers) and distribution function fits (curves) for (a)-(b) $s$-polarized optical pumping with photon energies of 2.05~eV and 2.52~eV and (c)-(d) circularly polarized optical pumping with a photon energy of 2.05~eV. The hole number was extracted using the analysis procedure in Fig. \ref{fig:3} for (a), (c) the upper VB and (b), (d) the lower VB. The calculated hole density $n_h$ per valley is provided in each panel. The insets of the WS$_2$ band diagrams present the possible transitions for the optical excitation conditions in each panel.}
  \label{fig:4}
\end{figure}

Results of the above analysis are compared for the various pump pulse settings in Fig. \ref{fig:4}, which shows the distribution of optically excited holes around the VBM. We fit the momentum dependence of $N_h$ with a hole-like Fermi-Dirac function $f_{h}(k) = 1 - f_{e}(k) = 1- 1/\{\exp[(\epsilon(k-k^{\prime})-\mu)/\delta]+1\}$ in the cases where $N_h > 0$ is clearly fulfilled. Here $\epsilon(k)$ is the bare dispersion of either the upper or lower VB, which is determined from the EDC fits of the band positions in Fig. \ref{fig:3}, and $\{k^{\prime},\mu,\delta\}$ constitute the set of fitting parameters that determine the position, height and width of the distribution. The fits of the Fermi-Dirac distribution show that the variation in $N_h(k)$ corresponds to a thermalized hole population for the bare band dispersion of the material. By assuming that the population of holes is isotropic around the VBM, we can convert the fitted distribution function into an estimate of the excited hole density per valley, $n_h$ \cite{Ulstrup:2016}. The detection limit for $n_h$ in this type of analysis is estimated to be 10$^{10}$ holes per cm$^2$. 

In Fig. \ref{fig:4}(a) we observe that $s$-polarized resonant pumping at the upper VB generates a clear distribution of holes with a density that is significantly larger than when we resonantly pump the lower VB, as seen in Fig. \ref{fig:4}(b). This general trend can be understood from the simple schematics of the excitation processes shown in the insets of Figs. \ref{fig:4}(a)-(b). However, the smaller hole density in the lower VB possibly also reflects a reduced lifetime of these hole states due to the increased phase space for scattering at the higher binding energies compared to the upper VB. Resonant pumping from the upper VB leads to free electrons and holes, which are situated at the band edges where the phase space for intra-band decay processes is much smaller, thereby making the optical pumping at 2.05~eV more efficient at generating free carriers. Pumping at this energy with circularly polarized pulses leads to a twice as large optically generated hole density in the upper band using $\sigma^{-}$- compared to $\sigma^{+}$-polarization, as seen in Fig. \ref{fig:4}(c). No significant hole density is detected in the lower band in Fig. \ref{fig:4}(d), which is consistent with the result for $s$-polarized pump pulses with the same energy in Fig. \ref{fig:4}(b). This analysis shows that, under these conditions, the free carriers excited in the $\bar{K}$-valley all derive from the upper band and must thus have a single spin quantum number.

Our results reveal that the level of control of the spin- and valley-degrees of freedom demonstrated for excitons in SL TMDCs also extends to optically generated free carriers in these materials. While our results show that these quantum numbers can be accessed by varying the photon energy and the polarization of a pump laser pulse for an epitaxial SL WS$_2$ crystal supported on a metallic substrate, we expect this scheme is transferable to arbitrary substrates and other SL TMDC materials.

We thank Phil Rice for technical support during the Artemis beamtime. We gratefully acknowledge funding from the VILLUM foundation, the Aarhus University Research Foundation, the Lundbeck foundation, EPSRC (Grant Nos. EP/I031014/1 and EP/L505079/1) and The Royal Society. Ph. H., J. A. M. and S. U. acknowledge financial support from the Danish Council for Independent Research, Natural Sciences under the Sapere Aude program (Grant Nos. DFF-4002-00029, DFF-6108-00409 and DFF-4090-00125). Access to the Artemis Facility was funded by STFC.

\end{document}